\documentclass[a4paper, amsfonts, amssymb, amsmath, reprint, showkeys, footinbib,twoside,superscriptaddress,floatfix,longbibliography]{revtex4-1}

\usepackage{graphicx} 
\usepackage{amsmath}
\usepackage{comment}
\usepackage{booktabs}
\usepackage{hyperref}
\usepackage{url}
\usepackage{adjustbox}
\usepackage{array}
\usepackage{xcolor}
\usepackage[USenglish]{babel}

\newcommand{\figLabelCapt}[1]{\textbf{\MakeLowercase{{#1}}}}
\newcommand{\refSub}[2]{\hyperref[#2]{\ref{#2}\figLabelCapt{#1}}}

\newcommand{\br}[1]{\mathbf{r}}
\newcommand{\bk}[1]{\mathbf{k}}

\begin{document}

\title{Machine learning interatomic potential can infer electrical response}

\author{Peichen Zhong}
\thanks{These authors contributed equally.}
\affiliation{Bakar Institute of Digital Materials for the Planet, UC Berkeley, California 94720, United States}

\author{Dongjin Kim}
\thanks{These authors contributed equally.}
\affiliation{Department of Chemistry, UC Berkeley, California 94720, United States}

\author{Daniel S. King}
\thanks{These authors contributed equally.}
\affiliation{Bakar Institute of Digital Materials for the Planet, UC Berkeley, California 94720, United States}

\author{Bingqing Cheng}
\email{bingqingcheng@berkeley.edu}
\affiliation{Bakar Institute of Digital Materials for the Planet, UC Berkeley, California 94720, United States}
\affiliation{Department of Chemistry, UC Berkeley, California 94720, United States}
\affiliation{The Institute of Science and Technology Austria, Am Campus 1, 3400 Klosterneuburg, Austria}

\date{\today}

\begin{abstract}
Modeling the response of material and chemical systems to electric fields remains a longstanding challenge. Machine learning interatomic potentials (MLIPs) offer an efficient and scalable alternative to quantum mechanical methods but do not by themselves incorporate electrical response. Here, we show that polarization and Born effective charge (BEC) tensors can be directly extracted from long-range MLIPs within the Latent Ewald Summation (LES) framework, solely by learning from energy and force data. Using this approach, we predict the infrared spectra of bulk water under zero or finite external electric fields, ionic conductivities of high-pressure superionic ice, and the phase transition and hysteresis in ferroelectric PbTiO$_3$ perovskite. This work thus extends the capability of MLIPs to predict electrical response--without training on charges or polarization or BECs--and enables accurate modeling of electric-field-driven processes in diverse systems at scale.
\end{abstract}

\maketitle

\section{Introduction}

The polarization $\mathbf{P}$ of a system underlies many electrical response properties 
including capacitance, dielectric constant, ferroelectricity, piezoelectricity, ionic conductivity, and infrared (IR) spectra.
The Born effective charge (BEC) tensor $Z^*$ quantifies the variation in $\mathbf{P}$ due to an atomic displacement at position $\mathbf{r}_i$ of atom $i$~\cite{wang2022dynamical}:
\begin{equation}
Z^*_{i \alpha\beta}  = \dfrac{\partial P_{\alpha}}{\partial r_{i\beta}}
= \dfrac{\partial \mathcal{F}_{i \alpha}}{\partial \mathcal{E}^0_\beta},
\label{eq:z}
\end{equation}
where $\alpha$ and $\beta$ label Cartesian directions. 
The second part of Eqn.~\eqref{eq:z} links the electrostatic force $\mathcal{F}_i$ on atom $i$ resulting from an external electric field $\boldsymbol{\mathcal{E}}^0$ to the system.

Modeling electrical response properties has long been a challenge. 
The Berry phase definition of the polarization~\cite{resta1994macroscopic,resta2007theory} of periodic insulators can be obtained from density functional theory (DFT) calculations. Density-functional perturbation theory (DFPT)~\cite{gonze1997dynamical} or the finite field method~\cite{zhang2020modelling} can be used to compute BECs and other derivatives of the polarization.
However, the computational costs associated with such ab initio methods limit their applications to large systems or long timescales.
On the other hand, fixed-charge or polarizable empirical force fields are cheap but
may lack quantitative accuracy or transferability~\cite{kirby2019charge,jorge2024theoretically}. 

Standard machine learning interatomic potentials (MLIPs)~\cite{keith2021combining,unke2021machine}, 
which learn surrogate potential energy surfaces from quantum mechanical reference calculations, are typically short-ranged and do not explicitly consider electrostatics.
Several approaches have been developed to incorporate long-range interactions,
such as learning DFT-derived partial charges~\cite{unke2019physnet,ko2021fourth,gong2024bamboo,shaidu2024incorporating} or 
maximally localized Wannier centers~\cite{zhang2022deep,gao2022self},
or employing long-range descriptors~\cite{grisafi2019incorporating,faller2024density} or long-range message-passing~\cite{kosmala2023ewald}.
Latent Ewald Summation (LES)~\cite{cheng2024latent,kim2024learning} is a recent method that learns a long-range energy contribution $E^\mathrm{lr}$ by fitting to the total potential energy $E$ and atomic forces $F$ of configurations:
\begin{equation}
E^\mathrm{lr} = \dfrac{1}{2\varepsilon_0V} \sum_{0<k<k_c} \dfrac{1}{k^2} e^{-\sigma^2 k^2/2} |S(\mathbf{k})|^2,
\label{eq:e_lr}
\end{equation}
where $\mathbf{k}$ is the reciprocal wave vector and $V$ is the cell volume. 
The structure factor $S(\mathbf{k})$ given by
\begin{equation}
S(\mathbf{k}) = \sum_{i=1}^N q_i^\mathrm{les} e^{i\mathbf{k}\cdot\mathbf{r}_i},
\label{eq:sfactor}
\end{equation}
and the LES charges $q_i^\mathrm{les}$ are predicted using a neural network based on local invariant features $B_i$ of atom $i$.
LES can be combined with any short-ranged MLIP architecture, such as descriptor-based~\cite{behler2007generalized,shapeev2016moment,drautz2019atomic} or message-passing neural networks~\cite{batzner20223,batatia2022mace}.

However, a natural inclusion of electric response directly within the MLIP frameworks is missing. 
Currently, $\mathbf{P}$ and BEC have to be learned separately as tensorial properties~\cite{grisafi2018symmetry,Rossi2025}, e.g. directly predicting BEC~\cite{Schmiedmayer2024} or local contribution to the dipole moment of a molecule~\cite{grisafi2018symmetry} based on atomic local environments,
or as the derivatives of the electric enthalpy~\cite{falletta2024unified}.
Conceptually, this is in contrast with the electronic structure picture of matter:
electron density and nuclear positions fully determine how the system will interact with electric field.

Here, we show that polarization and BEC tensors can be naturally derived from long-range MLIPs within the LES framework. 
This enables accurate predictions of electrical response properties, such as IR spectra and conductivity, solely by learning from energies and forces. 
Importantly, it is straightforward to add an external electric field to MLIP-driven molecular dynamics (MD) simulations, enabling the exploration of electric-field-driven phenomena in various materials and molecules. 
We demonstrate this method on a range of complex bulk systems, including molecular liquids, ionic liquids, superionic crystals, and ferroelectric materials.

\section{Theory}
Building on the pioneering Molecular Dynamics in Electronic Continuum (MDEC) model~\cite{leontyev2003continuum,leontyev2011accounting,kirby2019charge,jorge2024theoretically}, 
the Coulomb interactions between the ``free charges'' of atoms are explicitly considered, while the rapidly responding background electrons are treated as a dielectric medium. 
The electrostatic field produced by the free atomic charge $q_i$ of atom $i$ can then be expressed as
\begin{equation} 
\boldsymbol{\mathcal{E}}_{i} (\mathbf{r}) 
= \dfrac{q_i}{4\pi \varepsilon_0 \varepsilon_\infty |\mathbf{r}-\mathbf{r}_i|^3} (\mathbf{r}-\mathbf{r}_i), 
\end{equation} 
and the resultant electric force between two atoms is given by: 
\begin{equation} 
\mathcal{F}_{ij} = \boldsymbol{\mathcal{E}}_{i} (\mathbf{r}_{ij}) q_j, 
\end{equation} 
where $\varepsilon_0$ represents the vacuum permittivity, and $\varepsilon_\infty$ is the high-frequency relative permittivity (also known as the static or electronic dielectric constant), which can be determined experimentally (e.g., from the square of the optical refractive index) or calculated using DFPT with frozen nuclei~\cite{farahvash2018dynamic}.

We interpret the LES charges $q_i^\mathrm{les}$ as scaled charges,
$q_i^\mathrm{les} = q_i / \sqrt{\varepsilon_\infty}$,
acting both as the sources and the receivers of the electric field:
\begin{equation}
    \mathcal{F}_{ij} = \dfrac{q_i^\mathrm{les} q_j^\mathrm{les}}{ 4\pi \varepsilon_0 r_{ij}^3} \mathbf{r}_{ij}.
\end{equation}
In the LES algorithm, $q_i^\mathrm{les}$ are optimized to effectively describe the long-range component of the total potential energy and atomic forces.
For modeling the energetics and dynamics of a system under no external field, these LES charges provide a self-contained description of electrostatic interactions, with no further adjustment needed. 
To model electrical response, however, the LES charges can be unscaled to recover the atomic charges $q_i$.

The polarization of a finite system such as a gas-phase molecule is
\begin{equation}
    \mathbf{P}
    = \sum_{i=1}^N q_i \mathbf{r}_i ,
    \label{eq:P-finite}
\end{equation}
and the BEC $Z^*$ can be obtained by taking its derivative with respect to $\mathbf{r}$:
\begin{equation}
Z^*_{i \alpha\beta} 
= \dfrac{\partial P_{\alpha}}{\partial r_{i\beta}}
= q_i \delta_{\alpha,\beta}
+ \sum_{i=1}^N r_{j\alpha}\dfrac{\partial q_j}{\partial r_{i\beta}}.
\label{eq:z-finite}
\end{equation}
As shown in the second part of Eqn.~\eqref{eq:z-finite},  
the BEC tensor $Z_i^*$ comes from two contributions:
the charge $q_i$, and the dependence of the charges on atomic positions.

For modeling either crystalline or disordered bulk systems,
it is almost mandatory to apply periodic boundary (PBC) conditions.
Importantly, the value of $\mathbf{P}$ cannot be uniquely defined for
systems with PBC, according to the modern theory of polarization~\cite{resta1994macroscopic,resta2007theory}.
To circumvent such ambiguity, we propose a generalized formulation of polarization $P(k)$ under PBC:  
\begin{equation}
    P_\alpha(k) 
    = \sum_{i=1}^{N} \dfrac{q_i}{ik} \exp(i k r_{i\alpha}) ,
    \label{eq:P-complex}
\end{equation}
where $P_\alpha(k)$ is the polarization along the $\alpha$ direction, $k = 2\pi/L_\alpha$, and $L_\alpha$ is the length of a periodic cubic cell, while
the extension to triclinic cells is straightforward.
At the limit of $k \rightarrow 0$, 
$P(k)$ becomes the finite-system expression in Eqn.~\eqref{eq:P-finite}.
The BEC tensor $Z^*$ of atom $i$ can then be evaluated as
\begin{equation}
Z^*_{i \alpha\beta} 
= \Re\left[\exp(-ik r_{i\alpha}) \dfrac{\partial P_{\alpha} (k)}
{\partial r_{i\beta}}\right].
\label{eq:z-pbc}
\end{equation}

The prediction of BEC enables the calculation of several electrical response properties.
For instance, the current of the polarization of the system can be obtained as 
$\mathbf{J}(t) = \sum_{i=1}^N Z_i^*(t)\cdot\mathbf{v}_i(t)$.
The current-current autocorrelation function encodes the 
ionic electrical conductivity $\sigma$ via the Green-Kubo formula,
\begin{equation}
    \sigma = \dfrac{1}{3V k_B T} \int_0^\infty dt \left\langle \mathbf{J}(0) \mathbf{J}(t) \right\rangle,
    \label{eq:sigma}
\end{equation}
and the IR spectra via Fourier transform,
\begin{equation}
    I(\omega) \propto \int_0^T dt \left\langle \mathbf{J}(0) \mathbf{J}(t) \right\rangle e^{-i\omega t}.
    \label{eq:ir}
\end{equation}

Moreover, once BEC are computed using Eqn.~\eqref{eq:z-pbc}, one can apply a real-valued constant electric field $\boldsymbol{\mathcal{E}}^0$ to the system by adding the electrostatic force $\boldsymbol{\mathcal{F}}_i$ on each atom using the second part of Eqn.~\eqref{eq:z} for the linear response regime,
thus enabling constant-electric-field simulations under PBC.

\section{Examples}

\subsection{Water}

The theoretical prediction for the IR spectrum of water is a classic problem but still not fully resolved,
and more so with the presence of external
electric fields~\cite{cassone2019ab}.
We used the Cartesian Atomic Cluster Expansion (CACE)~\cite{cheng2024cartesian} potential as the short-ranged MLIP and LES~\cite{cheng2024latent,kim2024learning} as the long-range part, and thereafter refer to 
this combination as CACE-LR.
The RPBE-D3 bulk water dataset from Ref.~\cite{Schmiedmayer2024} contains energies and forces of 654 configurations (90\% train/ 10\% test split) each of 64 water molecules.
Even though we used a compact CACE model with a cutoff of 4.5~\AA{} and no message passing,
the test root mean square errors (RMSEs) of energy and forces 
(0.25~meV/atom and 21~meV/\AA) are a fraction compared to the errors in Ref.~\cite{Schmiedmayer2024} (0.8~meV/atom and 60~meV/\AA).

\begin{figure}
\centering
\includegraphics[width=\linewidth]{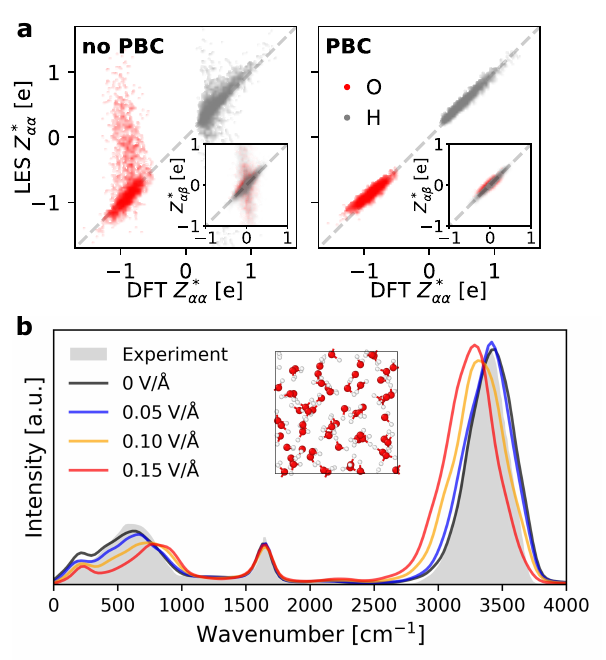}
     \caption{
     Electrical response of the RPBE-D3 bulk water.
     \figLabelCapt{a} compares the Born effective charge tensors ($Z^*$) computed from DFT and predicted using the LES method.
     The CACE-LR was trained on the energies and forces of the RPBE-D3 bulk water dataset~\cite{Schmiedmayer2024}.
     The main panels compare the diagonal elements of BEC ($Z^*_{\alpha\alpha}$), and the insets show the off-diagonal elements ($Z^*_{\alpha\beta}$ with $\alpha \ne \beta$).
     The left panel (no PBC) corresponds to the LES BECs calculated assuming no periodic boundary condition using Eqn.~\eqref{eq:z-finite},
     and the right panel (PBC) shows PBC obtained using the generalized polarization form in Eqns.~\eqref{eq:P-complex} and ~\eqref{eq:z-pbc}.
    \figLabelCapt{b} shows the infrared (IR) absorption spectra of bulk liquid water in the absence of an external field (black line) and under varying external field intensities (colored lines) as indicated in the legend. 
    The experimental IR spectrum in the absence of an external field ~\cite{Bertie1996Infrared} (gray shading) is included for reference.
    }
    \label{fig:water-bec}
\end{figure}

Fig.~\ref{fig:water-bec}a compares the BEC predicted by LES and calculated by the reference DFT for 100 water configurations at experimental density and room temperature~\cite{Schmiedmayer2024}.
If neglecting PBC and naively using Eqn.~\eqref{eq:z-finite}, the predicted BEC exhibited significant discrepancies from the DFT reference values, particularly for atoms near the edge of the simulation cell.
In contrast, by properly accounting for PBC using Eqn.~\eqref{eq:P-complex} and Eqn.~\eqref{eq:z-pbc}, the LES BECs agree well with DFT for both diagonal and off-diagonal components.
This shows that it is necessary to resolve the ambiguity of $\mathbf{P}$ for periodic systems using a scheme like Eqn.~\eqref{eq:P-complex}.

We performed MLIP-driven NVT simulations of bulk water (0.25 fs timestep and 200,000 MD steps) under the Nos\'e-Hoover thermostat, at 300~K and experimental density, and extracted the IR spectra using Eqn.~\eqref{eq:ir} based on the LES BECs.
As shown in Fig.~\ref{fig:water-bec}b (black curve), not only are the predicted shapes and positions of the intramolecular vibrational modes such as OH stretching band ($\approx 3400~\mathrm{cm}^{-1}$) and bending ($\approx 1640~\mathrm{cm}^{-1}$) mode band in excellent agreement with experiment~\cite{Bertie1996Infrared}, but also the intermolecular low-frequency libration mode band position~\cite{silvestrelli1997initio} ($\approx 650~\mathrm{cm}^{-1}$) and the hydrogen-bond translational stretching mode~\cite{heyden2010Dissecting} ($\approx 200~\mathrm{cm}^{-1}$) are well captured. 
Notably, 
it is necessary to use the time-dependent BEC tensors to compute the IR:
if using instead the fixed nominal charges of $q^H =+1$ for hydrogen and $q^O =-2$ for oxygen (see Fig.~\ref{fig:water_fixed_q_IR} in Methods), the shape of the predicted IR is much worse and the hydrogen-bond translational stretching band is completely absent. 

We then performed NVT MD simulations for bulk water under static constant external fields $\boldsymbol{\mathcal{E}}^0$ (0.05~V/\AA, 0.1~V/\AA, or 0.15~V/\AA) along the $z$-direction, 
by adding electrostatic forces on all atoms according to the second part of Eqn.~\eqref{eq:z}.
As shown in Fig.~\ref{fig:water-bec}b (colored curves), 
at higher electric field intensities, 
the intermolecular librational band ($\approx 650~\mathrm{cm}^{-1}$) blue shifts and the intramolecular OH stretching band ($\approx 3400~\mathrm{cm}^{-1}$) red shifts.
These trends are consistent with
previous studies using DFT molecular dynamics~\cite{cassone2019ab, futera2017Communication}.
The red-shift of the OH stretching band is generally associated with stronger hydrogen bonding~\cite{ojha2018Hydrogen} and with more ice-like structures~\cite{wang2004Vibrational}. 
The blue shift of the low-frequency libration mode can be attributed to the enhanced restrictions on the rotational motion of water molecules imposed by the H-bonds~\cite{cassone2019ab}.

\subsection{Superionic water}

\begin{figure}
\centering
\includegraphics[width=0.99\linewidth]{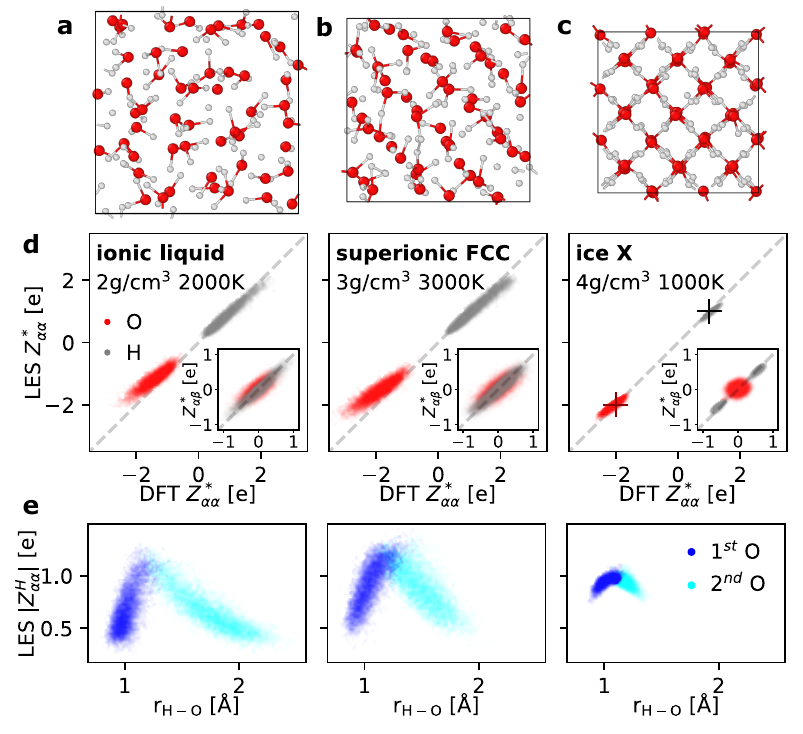}
     \caption{
     Analysis of the Born effective charges (BECs) in different phases of high-pressure water.
     \figLabelCapt{a} corresponds to partially ionic liquid water,
     \figLabelCapt{b} shows face-centred cubic (FCC) superionic phase (ice~XVIII),
     and \figLabelCapt{c} is ice X.
     The oxygen-hydrogen bonds are drawn with a cutoff of 1.1~\AA.
     \figLabelCapt{d} compares the BEC tensors ($Z^*$) computed from DFT and predicted using the LES method, for 100 configurations of each phase at the specified condition. 
     The CACE-LR was trained based on the energies and forces from the superionic water dataset~\cite{cheng2021phase}.
     The main panels compare the diagonal elements of BEC ($Z^*_{\alpha\alpha}$), and the insets show the off-diagonal elements ($Z^*_{\alpha\beta}$ with $\alpha \ne \beta$).
     \figLabelCapt{e} illustrates the correlation between the mean diagonal values of $Z^*$ of all hydrogen atoms, and the distances to their nearest two oxygen atoms.
    }
    \label{fig:superionic-bec}
\end{figure}

This example aims to illustrate the capability of our BEC inference method for ionic and superionic systems, and its generalizability across a wide range of conditions. Water at megabar pressures and thousands of Kelvins exhibits diverse structural and dynamical behaviors:
ionic water with partially dissociated hydrogen atoms (Fig.~\ref{fig:superionic-bec}a),
the face-centred cubic (FCC) superionic phase (ice~XVIII) with liquid-like hydrogen atoms while oxygen atoms remain on the crystalline lattice~\cite{Millot2019,cheng2021phase} (Fig.~\ref{fig:superionic-bec}b),
and ice X with a body-centered cubic (BCC) lattice of oxygen atoms~\cite{reinhardt2022thermodynamics} (Fig.~\ref{fig:superionic-bec}c).

We trained the CACE-LR model using 5,000 configurations (90\% train/ 10\% test split), randomly selected out of the 17,516 structures in the original MLIP training set that was compiled for predicting the phase diagram of superionic water spanning a wide range of thermodynamic conditions up to thousands of kelvin and megabar pressures~\cite{cheng2021phase}.
Despite the smaller training set and a relatively light-weight architecture choice of the CACE-LR model,
the test errors on energies and forces are halved compared to the original study: 
7 meV/atom in energy RMSE and 327 meV/\AA\ in force RMSE, compared to 14 meV/atom and 740 meV/\AA\ of the original MLIP~\cite{cheng2021phase}, respectively.

Fig.~\ref{fig:superionic-bec}b shows that the predicted LES BECs agree well with the ground-truth DFT values for three distinct phases (illustrated in Fig.~\ref{fig:superionic-bec}a-c) under different thermodynamic conditions.
At 2~g/cm$^3$ and 2000~K, the liquid water is partially molecular and partially dissociated with frequent
proton jumps.
The diagonal BEC values for hydrogen atoms ($Z^H_{\alpha\alpha}$) can sometimes exceed the nominal charge of $+1$. 
In Fig.~\ref{fig:superionic-bec}e, we correlate the distances between all H atoms and their nearest two oxygen atoms with the mean BEC diagonal values($3|Z^H_{\alpha\alpha}|^2 = (Z^H_{xx})^2 + (Z^H_{yy})^2 + (Z^H_{zz})^2$). 
This shows that the anomalously large BEC occurs when a hydrogen atom breaks the bond with its nearest oxygen and forms a bond with the second nearest oxygen, 
analogous to the Grotthuss mechanism.
At 3~g/cm$^3$ and 3000~K,
the FCC superionic ice exhibits even larger fluctuations of BECs for both hydrogen and oxygen.
The anomalous BECs of hydrogen are again correlated to the O-H bond breaking and formation,
and such events are more frequent.
At 4~g/cm$^3$ and 1000~K, the stable phase is ice X with 
a BCC lattice of oxygen atoms, and hydrogen atoms are evenly positioned between two neighboring oxygen atoms with straight O-H-O bonds. 
The diagonal values of BEC have very small fluctuations, and are centered around the oxidation numbers of $+1$ for hydrogen and $-2$ for oxygen ions (shown as crosses in the right panel of Fig.~\ref{fig:superionic-bec}d). 
Intriguingly, the off-diagonal elements of BEC for H show two separate clusters at positive and negative values.

\begin{figure}
\centering
\includegraphics[width=0.99\linewidth]{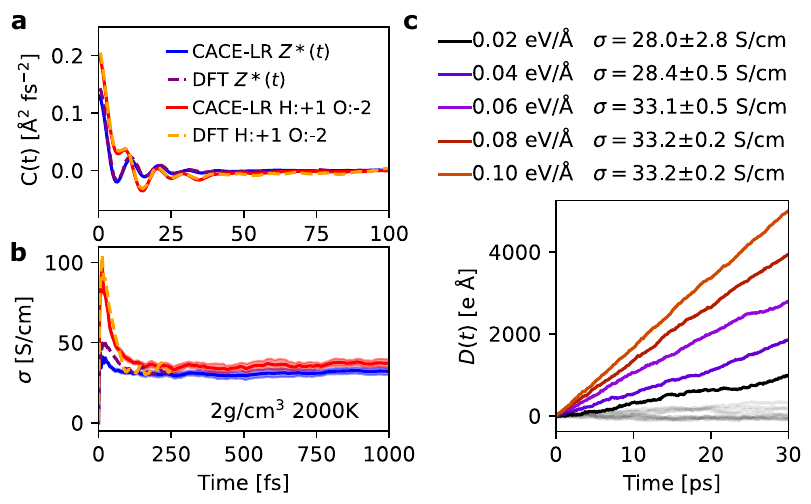}
     \caption{
     Ionic transport properties of the partially ionic liquid water at 2~g/cm$^3$ and 2000~K.
     \figLabelCapt{a} shows the current-current correlation functions $C(t)$ computed using either time-depedent Born effective charge tensors $Z^*(t)$ or fixed norminal charges.
     \figLabelCapt{b} plots the corresponding time integrals for estimating the ionic electrical conductivity $\sigma$.
     In \figLabelCapt{a} and \figLabelCapt{b}, the DFT molecular dyanmics results are from Ref.~\cite{french2011dynamical}.
     \figLabelCapt{c} illustrates the time-depedent charge displacement from CACE-LR molecular dynamics simulations under constant external electric fields with the specified intensities.
     The colored lines show the displacements along the direction of the applied field, and gray lines show the displacements along the other orthogonal directions.
    }
    \label{fig:superionic-sigma}
\end{figure}

The ionic electrical conductivity $\sigma$ is crucial for characterizing  
ionic and superionic systems.
To compute $\sigma$ at 2~g/cm$^3$ and 2000~K,
we employed a CACE-LR model that was finetuned using energy, forces, and BEC values of 100 configurations at the same condition.
We performed an equilibrium MD simulation of 120~ps duration with 54 water formula units, and such system size was selected to be directly comparable to the previous DFT MD simulation from Ref.~\cite{french2011dynamical}.
we calculated the current-current correlation functions $C(t)=\langle J(0)J(t)\rangle/3e$ either using the time-dependent BEC tensor $Z^*(t)$ or the fixed nominal charges of $q^H =+1$ and $q^O =-2$.
These correlation functions are shown in Fig.~\ref{fig:superionic-sigma}a,
and at short times they are in perfect agreement with a tour-de-force DFT MD and BEC calculation of 15~ps from Ref.~\cite{french2011dynamical}.
Fig.~\ref{fig:superionic-sigma}b shows the $\sigma$ values computed by integrating the corresponding $C(t)$ functions using the Helfand–Einstein formula~\cite{grasselli2021invariance}, a reduced-variance form of the Green-Kubo relation in Eqn.~\eqref{eq:sigma}.
Interestingly, the estimated $\sigma$ with time-dependent BEC tensors ($32\pm2$~S/cm) is similar to the estimate of
$\sigma=37\pm2$~S/cm with the constant $q^H = +1$ and $q^O = -2$.
Such observation was also made in Ref.~\cite{french2011dynamical}, and was rationalized in Ref.~\cite{grasselli2019topological} from a topological quantization argument albeit only for atomic liquids with all species adiabatically staying in
the same motifs without changing oxidation states.

One can also compute $\sigma$ from non-equilibrium MD simulations under external electric fields.
Fig.~\ref{fig:superionic-sigma}c shows the total displacement of charges in the system over time under different values of the external field $\mathcal{E}^0$ along the $z$ direction,
$D(t) = \sum_i^N q_i(z_i(t) - z_i(0))$, computed using the constant charges $q^H = +1$ and $q^O = -2$.
$\sigma$ can be estimated from the slope of $D(t)$ as 
$\sigma = \langle dD(t)/dt\rangle / V \mathcal{E}^0$.
Such $\sigma$ values from these finite-electric-field simulations, as displayed in the legend of Fig.~\ref{fig:superionic-sigma}c, are consistent with the equilibrium results,
but have better statistical convergence and also avoid the problems associated with the Green-Kubo integration to the infinite time limit (Eqn.~\eqref{eq:sigma}).

\subsection{Ferroelectric perovskite}

\begin{figure}[htb]
\centering
\includegraphics[width=\linewidth]{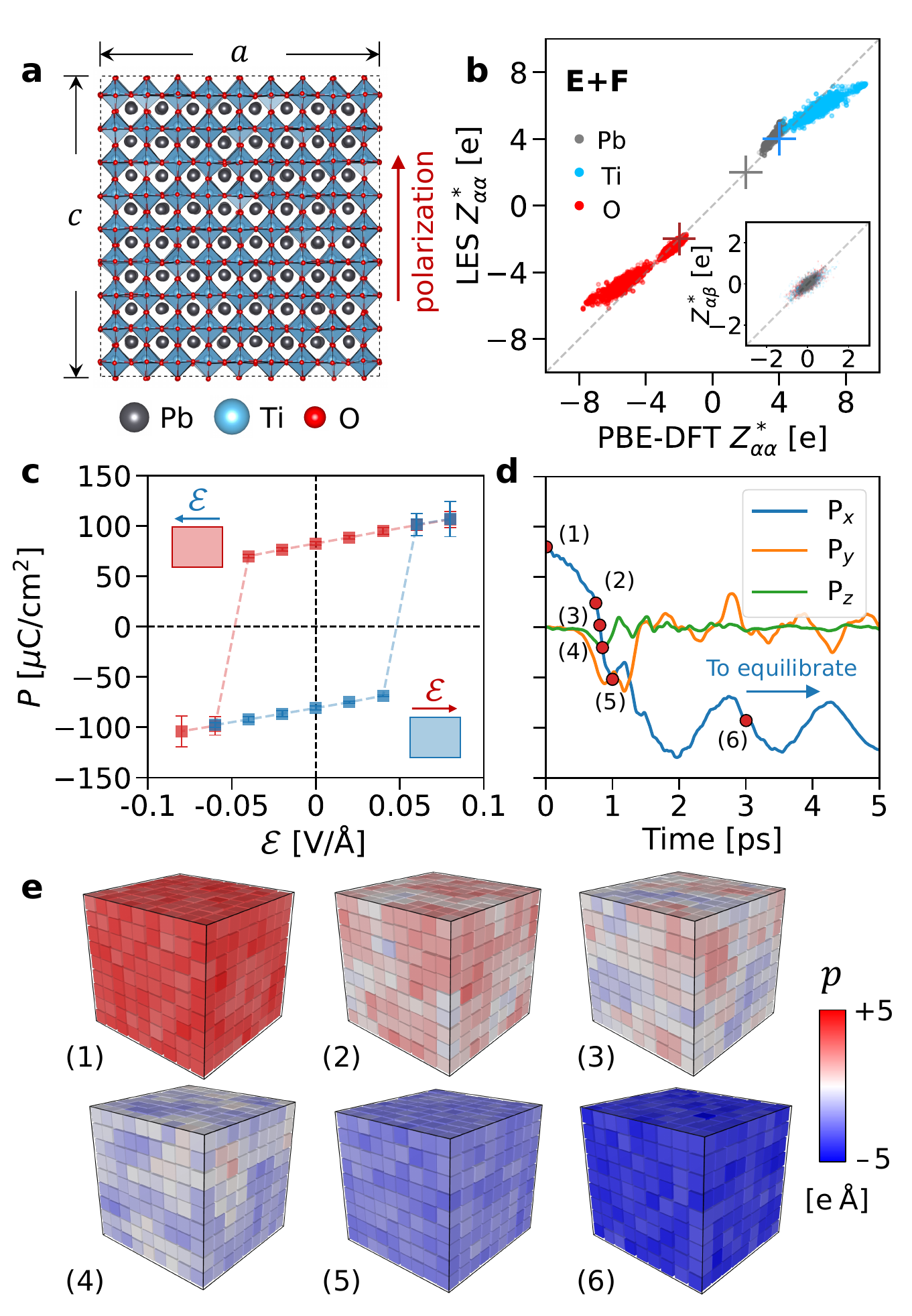}
\caption{
Polarization and Born effective charge tensors in the PbTiO$_3$ perovskite.
\figLabelCapt{a} A snapshot of equilibrated PbTiO$_3$ at $T=300$ K. 
\figLabelCapt{b} The BEC tensors ($Z^*$) computed from LES versus from PBE-DFT calculations for 45 randomly selected configurations. The $+$ symbol indicates the nominal charge of Pb/Ti/O in black/blue/red color.
\figLabelCapt{c} $P$-$\mathcal{E}$ hysteresis loop computed from CACE-LR MD simulations under different external electric fields.
\figLabelCapt{d} Time evolution of the total polarization during the non-equilibrium MD with $\mathcal{E} = -0.06$ V/\AA\ applied.
\figLabelCapt{e} Local dipole spatial distributions through the polarization reversal event. Each voxel represents a Ti-centered unit cell.}
\label{fig:PTO_polarization}
\end{figure}

Ferroelectric materials are unique in that they exhibit spontaneous and permanent electric polarization, and this polarization can be reversed by applying an electric field~\cite{rabe2007modern}.
Anomalously large BECs that exceed the nominal charges of ions are often considered hallmarks of ferroelectric materials~\cite{resta1995many,ghosez1998dynamical,filippetti2003strong}.
Here, we will demonstrate that our method can predict the anomalous BECs and model the characteristic $P$-$\mathcal{E}$ hysteresis loop in the prototypical PbTiO$_3$ ferroelectric perovskite.
At $T=300$ K, PbTiO$_3$ exhibits a tetragonal phase characterized by a short axis $a$ and a long axis $c$, with the ratio $c/a$ correlated with the polarization magnitude (see Fig.~\ref{fig:PTO_polarization}a).

We trained the CACE-LR model using the energies and forces of the PbTiO$_3$ dataset from Ref.~\cite{xie2024thermal}, computed using SCAN DFT. The potential achieved test RMSEs of 0.4 meV/atom in energy and 79.8 meV/\AA\ in forces, much reduced from the original model with 1 meV/atom in energy RMSE and 350 meV/\AA\ in force RMSE from Ref.~\cite{xie2024thermal}. 
Without explicitly learning the BEC, Fig.~\ref{fig:PTO_polarization}b compares the BEC tensors for 45 atomic configurations (including both cubic and tetragonal phase) predicted using LES and computed using PBE DFT.
Not only do the LES predictions agree well with the DFT reference, the anomalously large diagonal values of BECs ($Z^*_{\alpha\alpha}$) relative to the nominal ionic charges (indicated by the plus signs in Fig.~\ref{fig:PTO_polarization}b)
for $q^{\text{Pb}} = +2$, $q^{\text{Ti}} = +4$, and $q^{\text{O}} = -2$ are captured.
The anomalous BECs in ferroelectrics are typically considered to come from a complex interplay of global charge transfer, mixed ionic-covalent bonding, and the hybridization between oxygen and transition metal orbitals~\cite{ghosez1998dynamical}. The successful prediction of the BECs here thus not only showcases the expressiveness of the LES method, but also indicates that CACE-LR is able to capture the intricate long-range electrostatic interactions in ferroelectric materials~\cite{Zhong1994}. 
Unlike previous approaches~\cite{gigli2022thermodynamics,xie2024thermal,Rossi2025} that employ separate models for potential energy surfaces and polarization, CACE-LR properly embeds the latter into the former and do not explicitly train on polarization.

To characterize the spontaneous polarization of PbTiO$_3$ in the absence of an external electric field, we performed equilibrium NPT simulations at $T=300$~K (see Methods). 
Because SCAN-DFT overestimates $c/a=1.14$ for the ground state tetragonal PbTiO$_3$ structure compared to $c/a=1.06$ from experiments, we applied an isotropic external pressure of 2.8 GPa in the NPT simulations as used in Ref.~\cite{xie2024thermal}, yielding 
$c/a = 1.07$ for the equilibrated structures in MD.
We then calculated the total polarization of the material as~\cite{rabe2007first}:
\begin{equation}
    P_\alpha = \frac{e}{V}\sum_{i\beta} Z^{*}_{i\alpha\beta} \Delta u_{i\beta},
\end{equation}
where $\Delta u_{i\beta}$ indicates the atomic displacement from the non-polar centrosymmetric reference state.
The polarization predicted based on the LES BEC is $P = 82\pm 2$ $\mu$C/cm$^{2}$, consistent with the experimentally reported polarization $P(25^\circ\text{C})=81$ $\mu$C/cm$^2$~\cite{PhysRevB.7.3088_PTO_exp}.

We then simulated the $P$-$\mathcal{E}$ hysteresis loop of PbTiO$_3$ at $T=300$ K,
by applying an electric field $\mathcal{E}$ along and against the polarization direction of the equilibrated structure with a magnitude ranging from 0.02 to 0.08 V/\AA. 
Fig.~\ref{fig:PTO_polarization}c presents the time-averaged polarization at various external field strengths after equilibration, demonstrating polarization reversal to a negative value at $\mathcal{E}=-0.06$ V/\AA. Starting from the MD structure at $\mathcal{E}=-0.08$ V/\AA, we simulated the reverse polarization process by varying the electric field from $-0.06$ to $0.08$ V/\AA\ (blue line in Fig.~\ref{fig:PTO_polarization}c), which exhibits the expected reverse transition behavior and completes the ferroelectric hysteresis loop.
Note that the width of the hysteresis loop will be dependent on the system size and simulation time, as phase transition is an activated event, and the hysteresis loop in Fig.~\ref{fig:PTO_polarization}c aims to demonstrate the existence of hysteresis in the ferroelectric PbTiO$_3$.

Fig.~\ref{fig:PTO_polarization}d illustrates the time evolution of 
the total polarization during
the non-equilibrium MD trajectory after applying $\mathcal{E} = -0.06$ V/\AA\ to the original equilibrated structure at $T=300$ K. 
The polarization drops rapidly from its spontaneous value to negative values, followed by fluctuations before equilibrating to the new field-induced equilibrium state.
To visualize the atomic-level process for the polarization reversal, we calculated the local dipole moment $p$ for each Ti-centered unit cell using $p_{\alpha} = e\sum_j w_j Z^{*}_{j\alpha\beta} \Delta u_{j\beta}$,
where the sum runs over all atoms $j$ in the unit cell, and $w_j$ represents the weighting factor ($1$ for Ti, $1/8$ for Pb, $1/2$ for O).
Fig.~\ref{fig:PTO_polarization}e shows the spatial distributions of $p$ from representative MD snapshots, with the color of each voxel corresponding to the magnitude of $p$ along the $c$ axis.
Snapshots (2)--(4) reveal the initial stage of the reversal: domains of opposite polarization nucleate, grow, coalesce, and ultimately form a uniformly negatively polarized state.

\section{Discussion and conclusions}

The central thesis of the present paper is that MLIPs can infer the response of a material or chemical system to an external electric field,
by fitting latent charges ($q^\mathrm{les}$) just from the energies and forces of atomic configurations.
Although previous works provide some hints on such a connection, e.g. 
by showing that machine-learned charges are
correlated with DFT charges~\cite{song2024charge,kim2024learning},
this paper provides a clear physical picture and provides convincing demonstrations on diverse systems.

Our approach involves several conceptual advances. 
First, we recognize that LES charges are physical charges that can be determined by fitting to the energies and forces,
and that it is not necessary nor advantageous to explicitly fit to the semiclassical partial charges from DFT~\cite{kim2024learning}.
Indeed, the DFT partial charges are not physical observables and they depend on the specific partitioning strategy used~\cite{weinhold2016nbo,verstraelen2016minimal,marenich2012charge} and are less indicative of atomic charge states in oxides with charge transfer effects~\cite{Wolverton1998_LiCoO2}.
Our approach of not fitting to the DFT partial charges is in contrast with several other existing long-range MLIP methods~\cite{ko2021fourth,gong2024bamboo,zhang2022deep,shaidu2024incorporating}.

Second, we incorporate the background fast-responding charges into the high frequency relative permittivity $\epsilon_\infty$, while explicitly accounting for the Coulomb interactions between the atomic charges. 
This is similar to the concepts of screened Coulomb interactions and the scaled ion charge in the MDEC model~\cite{leontyev2003continuum,leontyev2011accounting}.
MDEC is foundational for designing and justifying non-polarizable force fields using scaled ion charges~\cite{kirby2019charge,jorge2024theoretically}, including the popular SPC-type water models~\cite{leontyev2010electronic,izvekov2004effective}.
The accuracy of the LES BEC based on $q_i= \sqrt{\varepsilon_\infty} q_i^\mathrm{les}$ provides a ``smoking-gun'' validation of the MDEC theory.
In addition, our approach is a cleaner way to use the charge scaling framework, as the flexible charges are learned from ab initio data.
In comparison,
the empirical force fields have the scaled charges and other parameters fitted at the same time to experimental data, which means the errors in describing the electrostatic interactions can be partially canceled by tuning other non-bonded parameters and vice versa,
so the resulting charges are less reflective of the true underlying electrostatics.
Moreover, the LES framework assigns flexible charges based on local atomic environments. 
Such environment-dependent charges are more expressive than the fixed charges in empirical force fields.
For example, the same LES model is capable of predicting BECs for dramatically different phases of water including isolating ice, superionic ice, and ionic water (Fig.~\ref{fig:superionic-bec}).
It is difficult to imagine a fixed-charge model to match this level of expressiveness.

Third, we derive the Born effective charge tensor $Z^*_i$ of each atom from the predicted unscaled charges $q$,
by taking the derivative of the total polarization $\mathbf{P}$ with respect to atomic positions. 
For periodic-boundary-condition systems, where $\mathbf{P}$ is not well-defined, we develop a generalized formulation (Eqn.~\eqref{eq:P-complex}). 
Unlike the DFT partial charge, which suffers from the lack of a unified definition, the BEC tensors are physical observables. 
The fact that the LES BECs align well with the DFT BECs proves that the LES charges are physical charges, despite not being trained on DFT partial charges. 
Moreover, the link between $q$ and $Z^*_i$ gives the option to train or finetune the MLIP using DFT BECs, e.g. as we did for the water, the superionic water and the PbTiO$_3$ system in the Methods section.
Successful prediction of BECs is also practically useful, as it can be used to predict a number of electrical response properties such as IR and conductivity. 
It also provides the linear response of forces to an applied electric field (Eqn.~\eqref{eq:z}), enabling straightforward incorporation of external fields in MLIP molecular dynamics simulations.
 
We demonstrated the framework on a diverse set of complex bulk systems, including liquid water, ionic high-pressure water, superionic water ice, insulating ice X, and ferroelectric PbTiO$_3$.
The LES predicted $Z^*$ are largely in good agreement with DFT, even when trained only on energies and forces, 
with further improvement possible by explicitly incorporating DFT BECs during the training. 
Notably, the model generalizes well across different phases and thermodynamic conditions, as seen in high-pressure water
(Fig.\ref{fig:superionic-bec}).
The framework also enables quantitative predictions of key electrical response properties,
such as the IR spectra of water, ionic conductivity in high-pressure water, and the $P$-$\mathcal{E}$ hysteresis loop in PbTiO$_3$.

To conclude, 
this paper resolves a critical limitation of state-of-the-art machine learning interatomic potentials: 
their inability to intrinsically predict electrical response properties. By rigorously linking the latent charges to the experimentally measurable BECs, we bridge the gap between MLIPs and the electrostatics in quantum mechanical systems. 
Our framework provides a systematic approach to develop and refine MLIPs for modeling polarizable systems under external fields, 
unlocking numerous applications such as electrolyte design, modeling electrochemical interfaces, piezoelectrics, and pyroelectrics.

\section{Methods}

\subsection{Water}

The RPBE-D3 bulk water dataset contains energies and forces of 654 configurations (64 water molecules in each snapshot), which were generated via an on-the-fly learning scheme from MD trajectories at different temperatures~\cite{Schmiedmayer2024}.
The original MLIP trained on this RPBE-D3 data from Ref.~\cite{Schmiedmayer2024} has a RMSE of 0.8 meV/atom and 60 meV/\AA{} for energies and forces, respectively.
Ref.~\cite{Schmiedmayer2024} also provided an additional RPBE-D3 dataset of 100 configurations that were separately collected from NVT simulations at experimental density and 298.2 K.
This set includes Born effective charges in addition to energies and forces.  
We will refer to this set as RPBE-D3 + BEC.

We trained four versions of CACE-LR: (1) trained solely on energies and forces with a cutoff radius $r_\mathrm{cut}=5.5$~\AA{} using RPBE-D3 data,
(2) trained solely on energies and forces with $r_\mathrm{cut}=4.5$~\AA{} using RPBE-D3 data,
(3) trained solely on energies and forces with a smaller dataset and $r_\mathrm{cut}=4.5$~\AA{} using RPBE-D3 + BEC data, and
(4) trained on energies, forces, and Born effective charges with a smaller dataset and $r_\mathrm{cut}=4.5$~\AA{} using RPBE-D3 + BEC data.
For the CACE representation, we used 6 trainable Bessel radial functions, $c = 12$, $l_\mathrm{max} = 3$, $\nu_\mathrm{max} = 3$, $N_\mathrm{embedding} = 2$, no message passing, 1-dimensional hidden variable, $\sigma = 1$~\AA{}, and $k_c = \pi$ ($dl=2$~\AA).

Table.~\ref{tab:4pots} summarizes the number of configurations in each data set used for training, along with the applied cutoff settings and the corresponding RMSE values for energies and forces.

DFPT calculations with the RPBE-D3 functional using VASP predict the high-frequency permittivity ($\varepsilon_\infty$) of water at experimental density and room temperature to be 1.83. We use this value when converting LES charges $q^\mathrm{les}$ to the physical charges $q$.
Such value is very close to the experimental value of 1.78 for water (with the refractive index of
water of 1.333 being the square root of this value).

\begin{figure}
\centering
\includegraphics[width=0.6\linewidth]{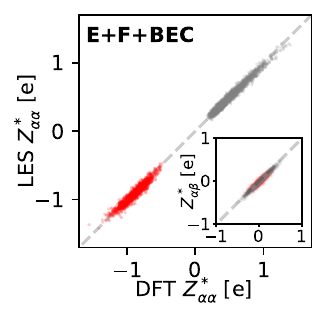}
     \caption{
     Comparison of DFT BEC and LES BEC using the version (4) potential that is trained 100 structures with energy, forces, and BEC.
    }
    \label{fig:water-bec-trained}
\end{figure}

Fig.~\ref{fig:water-bec-trained} compares the reference DFT and LES BEC predicted by the model trained on BEC (version 4) for the same 100 water configurations used in Fig.~\ref{fig:water-bec}. As expected, training directly on BEC data improves agreement between LES and reference DFT results, although the improvement is modest. This indicates that while prediction accuracy can be further enhanced by training with BEC data, the potential trained exclusively on energies and forces (version 2, Fig.~\ref{fig:water-bec}) already exhibits sufficiently high accuracy.
In the main text, all the reported results are from the version (2).

\begin{table}[h!]
\centering
\begin{tabular}{|l|c|c|c|c|}
\hline
Version & 1 & 2 & 3 & 4 \\
{} &  E+F &  E+F &  E+F  &  E+F+BEC \\
$N_{config}$ &  654 &  654 &  100  &  100 \\
$r_\mathrm{cut}$(\AA{}) &  5.5 & 4.5 & 4.5 & 4.5 \\
\hline
E &         0.22 &         0.25 &          0.26 &          0.19 \\
\hline
F &        18.88 &        21.01 &         23.84 &         25.34 \\
\hline
\end{tabular}
\caption{Performance of four versions of the CACE-LR potentials on each test set. Errors are reported via RMSE in meV/atom for energy and in meV/\AA{} for forces.}
\label{tab:4pots}
\end{table}

\begin{figure}
\centering
\includegraphics[width=0.99\linewidth]{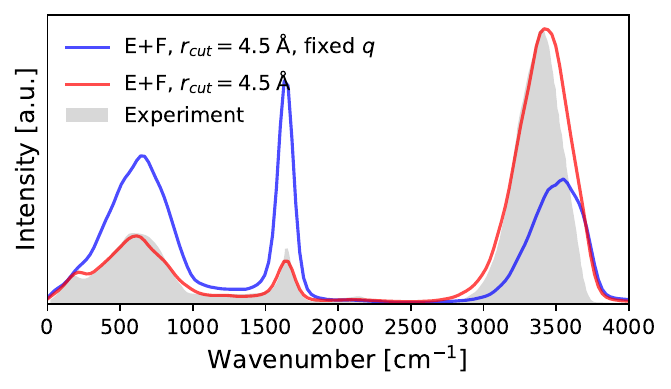}
     \caption{
     IR spectrum computed using fixed nominal charges ($q^H = +1$ for hydrogen and $q^O = -2$ for oxygen) based on the MD trajectory generated with model version (2). The experimental IR spectrum~\cite{Bertie1996Infrared} is included as gray shading for reference.
    }
    \label{fig:water_fixed_q_IR}
\end{figure}

We performed equilibrium NVT simulations in ASE at a density of 0.997~g/cm$^{3}$ and 300~K for a system of 64 water molecules, employing the Nos\'e-Hoover thermostat. 
The finite-field MD simulations followed the same setups. 
As the sum of $Z^*$ is not exactly zero due to the small residual prediction errors, in the finite-field MD simulations
the total mean forces on all atoms were removed every step to eliminate the non-zero center-of-mass velocities arising under the electric field.
Although such mean forces do not affect any physical observables, they can interfere with the thermostat and the visualization of the trajectories.
In all cases, MD simulations were conducted for 50~ps with a time-step of 0.25~fs. IR spectra were calculated from the MD trajectories employing Eqn.~\eqref{eq:ir}, which involves computing the polarization current of the system, \(\mathbf{J}(t) = \sum_{i=1}^{N} Z_i^*(t) \cdot \mathbf{v}_i(t)\). A Gaussian filter was applied following the Fourier transform, and each IR spectrum was normalized by its integrated area.

\begin{figure}
\centering
\includegraphics[width=0.99\linewidth]{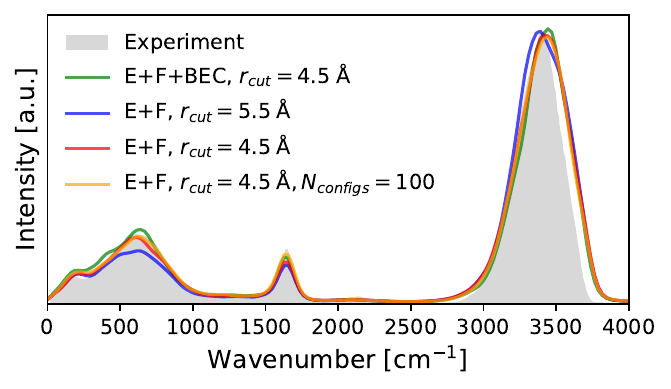}
     \caption{
     Water IR spectrum with four differently trained MLIP.
     Comparison of computational IR spectra of liquid water obtained from four differently trained MLIPs based on RPBE-D3 data. 
     The experimental IR spectrum~\cite{Bertie1996Infrared} is included as gray shading for reference.
    }
    \label{fig:water-IR-comparison}
\end{figure}

Fig.~\ref{fig:water_fixed_q_IR} shows that the IR spectrum calculated using fixed nominal charges does not reproduce peak intensities and shapes well. Moreover, it completely fails to capture the hydrogen-bond stretching mode at approximately $200~\mathrm{cm}^{-1}$. These results highlight the importance of accurately predicting Born effective charges to obtain detailed and reliable IR spectra.

Fig.~\ref{fig:water-IR-comparison} presents computational IR spectra of liquid water obtained using four differently trained potentials (Table.~\ref{tab:4pots}). All four versions of the CACE-LR potentials exhibit consistent peak positions and intensity trends, closely matching the experimental IR spectrum. Notably, these potentials differ in training dataset size, cutoff radius, and inclusion of Born effective charges, demonstrating robustness and reliability across varying training conditions.

\subsection{Superionic water}

The original training set of superionic water has 17,516 configurations  (90\% train/10\% test split), spanning a wide range of thermodynamic conditions (300~K-15000~K, 1~g/cm$^{3}$-7~g/cm$^{3}$),
and it was trained using N2P2~\cite{singraber2019parallel} with a cutoff of 12~Bohr,
yielding test RMSE errors of 14~meV/atom in energy, and 740~meV/A in forces~\cite{cheng2021phase}.

We randomly selected 5,000 configurations (90\% train/10\% test split) from the original dataset.
For the CACE-SR part, we used $r_\mathrm{cut} = 3.5$~\AA, 6 Bessel radial functions, $c = 12$, $l_\mathrm{max} = 3$, $\nu_\mathrm{max} = 3$, $N_\mathrm{embedding} = 3$, and 1 message passing layer.
The LES model uses a one-dimensional hidden variable, $\sigma = 1$~\AA{}, and $k_c = \pi$ ($dl=2$~\AA). 
The test RMSEs are 7 meV/atom in energy, and 327 meV/A in forces.

For comparing BECs,
we randomly selected 100 configurations of 54 water molecules at 3~g/cm$^3$ and 3000~K from DFT MD trajectories from Ref.~\cite{french2011dynamical}.
We also selected 100 uncorrelated configurations of 54 water molecules from MLP MD simulations at 2~g/cm$^3$ and 2000~K,
and at 4~g/cm$^3$ and 1000~K.
We employed VASP to calculate the Born effective
charge tensor for these configurations using DFPT, with a plane-wave cutoff of 400~eV at the
Baldereschi point, consistent with Ref.~\cite{french2011dynamical}.
From DFPT, we also computed the high-frequency relative permittivity at the three conditions:
$\varepsilon_\infty=3.1$ at 2~g/cm$^3$ and 2000~K,
$\varepsilon_\infty=4.2$ at 3~g/cm$^3$ and 3000~K,
and $\varepsilon_\infty=3.7$ at 4~g/cm$^3$ and 1000~K.
These values were used to infer the BECs.

We further finetuned the CACE-LR model 
using the energy, forces, and BEC values of the 100 configurations (90\% train/10\% test split) at 2~g/cm$^3$ and 2000~K.
Before the finetuning, the RMSE errors in energy, forces, and BEC are 4.5~meV/atom, 103~meV/A, and 0.136~e, respectively.
After the finetuning, the test RMSE errors reduced to 0.67~meV/atom and 101~meV/A, and 0.09~e, respectively.

For computing conductivities, we used this finetuned model to perform equilibrium NVT simulations in ASE at 2~g/cm$^3$ and 2000~K for a system of water molecules, employing the Nos\'e-Hoover thermostat.
The timestep was set to 0.3~fs, and the simulation length is 120~ps (400,000 steps) following 3~ps of equilibration.
The finite-field MD simulations follow a similar setup, except that a shorter simulation time of 30~ps (100,000 steps) was used.

\subsection{PbTiO$_3$ perovskite}

To fit the CACE-LR potential, we used the original training (4432 configurations) and test datasets (600 configurations) from Ref.~\cite{xie2024thermal} and randomly allocated 10\% of the original training data as the validation set. Both the training and test dataset contain the SCAN-DFT calculated energy and forces of PbTiO$_3$ atomic configurations from DP-GEN MD simulations at 300/600/900 K, covering the cubic (ferroelectric) and tetragonal (paraelectric) phases in $3\times3\times3$ PbTiO$_3$ unit cells (135 atoms).
For the CACE-SR part, we used $r_\mathrm{cut} = 6.0$~\AA, 6 Bessel radial functions, $c = 12$, $l_\mathrm{max} = 3$, $\nu_\mathrm{max} = 3$, $N_\mathrm{embedding} = 3$, and 1 message passing layer.
The LES model uses a one-dimensional hidden variable, $\sigma = 1$~\AA{}, and $k_c = \pi$ ($dl=2$~\AA).
The CACE-LR potential trained only on SCAN DFT energy and forces denotes $E+F$ model. 

\begin{figure}
\centering
\includegraphics[width=0.6\linewidth]{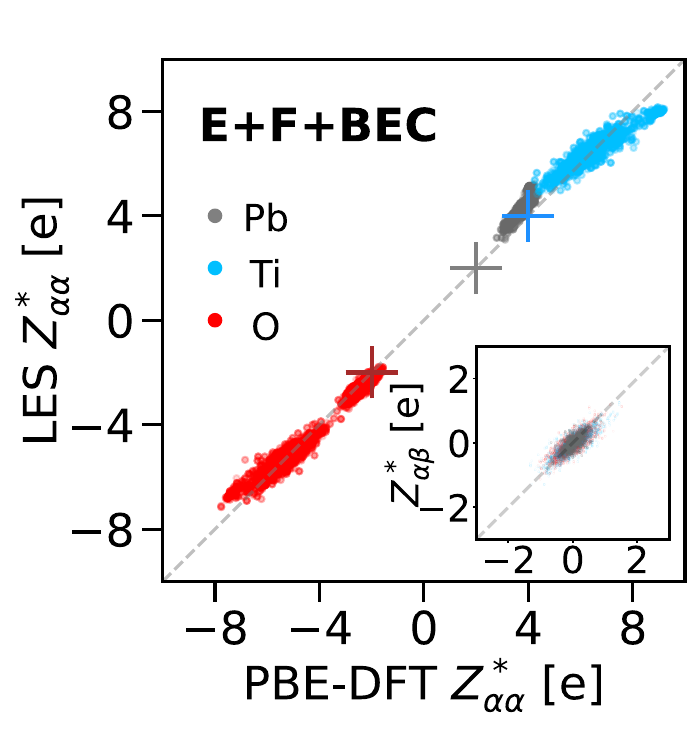}
     \caption{
     Comparison of DFT BEC and LES BEC using the potential that is trained 354 structures with energy, forces, and BEC.
    }
    \label{fig:PTO-bec-trained}
\end{figure}

To evaluate the LES BECs against DFT BECs, we randomly selected 443 atomic configurations from the training set for DFT calculations. These configurations were further split into train/val/test sets with a ratio of 8:1:1 for fine-tuning purposes. The test set comprised 45 atomic configurations and was used for comparison in Fig.~\ref{fig:PTO_polarization}.
For computing LES BECs, we replicated the supercells of these configurations $3\times3\times3$, to eliminate the finite-size effects due to finite $k$.
As a proof of concept, the DFT-BEC was calculated with DFPT at the generalized gradient approximation (GGA) level of accuracy, as the SCAN functional is not currently supported for DFPT. According to our test, the LEC-BEC derived from SCAN-DFT demonstrates good agreement with the PBE-DFT BEC.
The $E+F$ model was further fine-tuned with the PBE-DFT energy, forces, and BEC. 
Fig.~\ref{fig:PTO-bec-trained} displays the comparison after fine-tuning, which demonstrates a marginal improvement from an RMSE of 0.586 e to 0.384 e for diagonal components and from 0.257 e to 0.254 e for off-diagonal components of the BEC tensors when comparing the $E+F$ model to the fine-tuned model.
The DFPT calculations were performed using VASP with the PBE54 functional~\cite{perdew1996_GGA}, a Gamma-centered $k$-point, and a plane-wave energy cutoff of 680 eV. The DFT calculations were performed on $3\times3\times3$ the PbTiO$_3$ unit cell, with convergence of $10^{-6}$ eV in total energy.

For computing the ferroelectric properties of PbTiO$_3$, we employed the $E+F$ model to perform equilibrium NPT simulations using ASE at $P=P_0 + P_a$ with the Nos\'e-Hoover thermostat. Here, $P_0$ represents the ambient pressure (1 bar) and $P_a= 2.8$ GPa is an applied correction to compensate for the DFT overestimation of the $c/a$ ratio as suggested in Ref.~\cite{PhysRevLett.73.1861_Pa}. The simulation structure was initialized with a $9\times9\times9$ supercell of the cubic PbTiO$_3$ unit cell (space group $Pm3\Bar{m}$).
The MD simulations were conducted with a timestep of 2~fs. For simulations without external electric fields, we performed 100~ps production runs following 10~ps of equilibration.
The finite-field MD simulations followed a similar protocol, except that a shorter simulation time of 50~ps was used.

To estimate the high-frequency dielectric constant $\varepsilon_\infty$, we additionally performed MD simulations with a $3\times3\times3$ supercell at $T=300$ K to sample equilibrated structures from the MD trajectories. These atomic configurations were subsequently analyzed using DFPT to obtain the microscopic dielectric constant. We calculated the averaged diagonal value $\varepsilon_\infty = 1/N_{i\alpha}\sum_{i\alpha}\varepsilon_{\alpha\alpha}=7.533$ of the dielectric constant tensor, which was then used as the scaling factor of $\sqrt{\varepsilon_\infty}/9.48933$ to compute the polarization of PbTiO$_3$ at large scales from MD simulations. For the plots in Fig.~\ref{fig:PTO_polarization}b-e, a parity transformation of $-1$ was applied to align the BEC and polarization direction with conventional notation.

\subsection{Notes on implentation}
The LES method was implemented in the CACE code,
\url{https://github.com/BingqingCheng/cace}.
In practice, we use an \texttt{Atomwise} module to predict an internal hidden charge $q_i^\mathrm{raw}= Q_{\phi}(B_i)$ based on a set of local invariant representations $B_i$. The long-range energy is then computed using an \texttt{Ewald} module as
\begin{equation}
E^\mathrm{lr} = \dfrac{2\pi}{V} \sum_{0<k<k_c} \dfrac{1}{k^2} e^{-\sigma^2 k^2/2} |\sum_{i=1}^N q_i^\mathrm{raw} e^{i\mathbf{k}\cdot\mathbf{r}_i}|^2.
\end{equation}

To obtain the LES charges $q^\mathrm{les}$ in the unit of [e], 
the internal hidden charges $q^\mathrm{raw}$ should be scaled by a factor of $1/9.48933$, due to the internal normalization factor used ($1/2\varepsilon_0=1$).

We then use a \texttt{Polarization} module to compute the polarization of the system based on $q_i^\mathrm{raw}$. If the system is finite, the non-periodic expression (Eqn.~\eqref{eq:P-finite}) is used,
and if the system is periodic, the generalized polarization in Eqn.~\eqref{eq:P-complex} is used.
One can add a normalization factor in this module.
The default setting is to remove the mean average charge before computing the polarization.
If the factor of $\sqrt{\epsilon_{\infty}}/9.48933$ is used, the correct magnitude of the polarization will be recovered.

We then use the \texttt{Grad} module to take the derivative of polarization with respect to atomic positions using autograd (see Eqn.~\eqref{eq:z-finite}).
For finite systems, this step already provides the BECs.
For periodic systems, however, we need to use the \texttt{Dephase} module to remove the complex phase factor in Eqn.~\eqref{eq:z-pbc}, in order to get the real-valued BECs.

\textbf{Data availability}~The training sets, training scripts, and trained CACE potentials are available at \url{https://github.com/BingqingCheng/LES-BEC}.

\textbf{Code availability}~The CACE package is publicly available at \url{https://github.com/BingqingCheng/cace}.

\textbf{Acknowledgments}~
The authors thank for valuable discussions with Pinchen Xie, David Limmer, Jeff Neaton and Greg Voth.
The authors thank Sebastien Hamel for providing the DFT MD trajectories for superionic water, and help clarifying questions related to the pseudopotentials.
The authors thank Federico Grasselli and Stefano Baroni for providing data and notebooks for computing the conductivity of a molten salt.
D.S.K. and P.Z. acknowledge funding from BIDMaP Postdoctoral Fellowship.

\textbf{Competing Interests}~
B.C. has an equity stake in AIMATX Inc.
University of California, Berkeley has filed a provisional patent for the Latent Ewald Summation algorithm.

\end{document}